\documentclass[letterpaper, 10 pt, conference]{ieeeconf}
\IEEEoverridecommandlockouts                              
\usepackage{cite}
\usepackage{graphicx}      
\usepackage{amsmath}
\usepackage{amssymb}
\usepackage{subcaption}

\usepackage{amsthm}
\usepackage{mathrsfs}
\usepackage{empheq}
\usepackage{tikz}
\usepackage[utf8]{inputenc}
\usepackage{pgfplots} 
\usepackage{pgfgantt}
\usepackage{pdflscape}
\pgfplotsset{compat=newest} 
\pgfplotsset{plot coordinates/math parser=false}
\usepackage{tikzscale}
\usepackage{standalone}
\usetikzlibrary{shapes,arrows}
\usetikzlibrary{arrows,calc,positioning,fit}

\usepackage{stfloats}

\newtheorem{theorem}{Theorem}
\newtheorem{definition}{Definition}

\newtheorem{lemma}{Lemma}

\newtheorem{example}{Example}

\newcommand{\norm}[1]{\ensuremath\left\lVert{#1}\right\rVert}

\title{\LARGE \bf Characterizing nonlinear systems with mixed input-output properties through dissipation inequalities}
\author{Sebastiaan van den Eijnden and Thomas Chaffey
\thanks{Sebastiaan van den Eijnden is with the Department
of Mechanical Engineering, Eindhoven University of Technology, 5600 MB Eindhoven, The Netherlands.  E-mail: s.j.a.m.v.d.eijnden@tue.nl. \newline \indent
Thomas Chaffey is with the Department of Engineering, University of Cambridge, Trumpington Street, Cambridge CB2 IPZ.  His research is supported by Pembroke College, Cambridge.}}

\begin{document}

\maketitle
\pagestyle{empty}
\thispagestyle{empty}

\begin{abstract} 
Systems that show different characteristics, such as finite-gain and passivity, depending on the nature of the inputs, are said to possess \emph{mixed} input-output properties. In this paper, we provide a constructive method for characterizing mixed input-output properties of nonlinear systems using a dissipativity framework. Our results take inspiration from the generalized Kalman-Yakubovich-Popov lemma, and show that a system is mixed if it is dissipative with respect to highly specialized supply rates. The mixed input-output characterization is used for assessing stability of feedback interconnections in which the feedback components violate conditions of classical results such as the small-gain and passivity theorem. We showcase applicability of our results through various examples. 
\end{abstract}

\vspace{-.03cm}
\section{Introduction}
The small-gain and passivity theorem are undoubtedly two of the most fundamental results in input-output stability theory for feedback systems \cite{Zames66, Willems71, Desoer09}. The small-gain theorem guarantees stability of the negative feedback interconnection of two stable systems provided the product of their gains is less than one. The passivity theorem guarantees stability if, for example, one of the systems is passive, and the other one is strictly passive \cite{Zames66, Willems71, Desoer09, Khalil02, Teel96}. The strength of these results comes from the fact that they provide a generic method for anticipating the qualitative behaviour of a feedback interconnection with only rough information about the properties of the feedback components \cite{Teel96}. 

At the same time, this generality makes the small-gain and passivity theorems conservative. All input-output information is ``lumped'' into a single system property (small-gain or passivity) that may not always accurately reflect the complete system behavior. Consequently, there are many feedback systems which are stable, but do not meet the assumptions of the passivity theorem nor the small-gain theorem. For instance, flexible robotic manipulators would be passive systems, were it not for high frequency dynamics destroying passivity due to the presence of actuators and sensors \cite{Forbes13}. The input-output gain, however, is small at these high frequencies. In fact, many (if not all) mechanical structures naturally possess this type of combined passive and small-gain behaviour, and thus it is a natural way of characterizing input-output behavior of (nonlinear) systems. 

The prevalence of systems with combined passivity and small-gain characteristics has led to several specialized stability results that aim at relaxing the conditions of the classical small-gain and passivity theorems. For example, so-called \emph{mixed small-gain/passive} systems are considered in \cite{Griggs07, Griggs09, Griggs11, Forbes10, Forbes13, Forbes13_2, Xiao22}. These works characterize mixed properties through a pair of linear operators to define a “blended” supply rate. More recently, the work in \cite{Chaffey22} characterizes mixed properties in terms of the scaled relative graph. The idea takes inspiration from the blended supply rate in \cite{Griggs09}, however, rather than using a smoothly blended supply, the space of input signals is split into those pairs of signals where the two systems have small-gain, are passive, or both. Besides small-gain and passivity properties, combinations of other properties including small-gain, passivity, and negative imaginaryness have been considered \cite{Patra11, Das15}. 

All of the aforementioned works characterize mixed system properties in an input-output setting, but the link with Lyapunov/storage functions and the framework of dissipative dynamical systems \cite{Willems72} is missing. This is somewhat surprising as for classical small-gain and passivity results such links are well established \cite{vdSchaft17}. The main contribution of this paper is to provide a connection between mixed input-output properties and the framework of dissipative dynamical systems. We show that a system is mixed, if it is dissipative with respect to highly specialized supply rates. Our result is inspired by the generalized Kalman-Yakubovich-Popov (KYP) lemma for LTI systems \cite{Iwasaki, Iwasaki05, Pipeleers}. To the best of the authors' knowledge, a link between mixed properties for nonlinear systems and the generalized KYP lemma has not been established before.  

The results in this paper provide a different avenue for determining whether a nonlinear system possesses mixed input-output properties, and in that sense enrich the classical small-gain, passivity, and more general dissipativity results. Moreover, we believe our results to be valuable in the context of the recently developed framework of scaled relative graphs \cite{Chaffey23, Chaffey22}. Scaled relative graphs provide an elegant means for graphically verifying (robust) stability margins and performance for nonlinear systems, but estimating the scaled graph of a nonlinear system has been a difficult task so far, as this requires characterizing input-output behaviour for an infinite number of inputs. The same goes for determining whether or not a nonlinear system is ``mixed'' \cite{Griggs09}. The ideas put forward in this paper shed a new light on how to tackle these important problems (see also our related work in \cite{Eijnden24}). 

The remainder of this paper is organized as follows. Section~\ref{sec:prob} introduces the problem statement and discusses some preliminary results that serve as the main inspiration for the approaches taken in this paper. The main results are presented in Section~\ref{sec:main} in the form of a dissipation-based characterization of mixed system properties. We subsequently use these characterizations to formulate an interconnection result, and provide illustrative examples. Conclusions and an outlook for future work are presented in Section~\ref{sec:conclusions}.

\textbf{Notation.} The sets of $n$-by-$n$ symmetric matrices are denoted by $\mathbb{S}^{n}=\{P\in\mathbb{R}^{n\times n}\mid P=P^\top\}$. For $P \in \mathbb{S}^n$, we use $P_{ij}$ to indicate the $(i,j)$-th element of $P$, and $P\succ 0$ and $P\prec 0$ mean, respectively, that $P$ is positive definite, i.e., $x^\top P x>0$ for all $x\in\mathbb{R}^{n}\setminus\{0\}$, and negative definite, i.e., $-P\succ 0$. By $\|P\|_2$ we mean the spectral norm of a matrix $P$, that is, $\|P\|_2 = \sqrt{\lambda_{\textup{max}}(P^\top P)}$, where $\lambda_\textup{max}$ denotes the largest eigenvalue. For signals $u, y: [0,T] \to \mathbb{R}^n$ we denote 
\begin{equation*}\label{eq:int}
\langle u,y\rangle_T := \int_{0}^T u(t)^\top y(t) dt, \:\:\textup{ and } \:\: \|u\|_T = \sqrt{\langle u,u\rangle_T}.
\end{equation*} 
For $T = \infty$ we adopt the standard notation $\langle u,y \rangle_\infty = \langle u,y\rangle$, and $\|u\|_\infty = \|u\|$. 
The space of signals which are square-integrable over any finite time interval $[0, T]$, i.e., $\|u\|_T < \infty$, is denoted by $\mathcal{L}_{2e}$. We let $\mathcal{L}_2$ denote the space of square integrable signals on the time axis $[0, \infty)$. 

\section{Problem statement}\label{sec:prob}
In this section we introduce the system setting and problem statement, and discuss the generalized KYP lemma for LTI systems, which serves as the main inspiration for our results.  
\subsection{System setting}
In this paper, we consider nonlinear systems of the form
\begin{equation}\label{eq:nl}
    \begin{split}
\dot{x}(t) &= f(x(t),u(t)), \quad x(0)=0,\\
y(t) &= g(x(t),u(t)),
    \end{split}
\end{equation}
with states $x(t)\in \mathbb{R}^n$, input $u(t) \in \mathbb{R}$, and output $y(t) \in \mathbb{R}$ all at time $t \in \mathbb{R}_{\geq 0}$. Furthermore, $f:\mathbb{R}^n \times \mathbb{R} \to \mathbb{R}^n$ and $g:\mathbb{R}^n \times \mathbb{R} \to \mathbb{R}$ are nonlinear functions with $f(0,0)= 0$, and $g(0,0)=0$. Solutions to \eqref{eq:nl} are considered as absolutely continuous functions $x : [0,T] \to \mathbb{R}^n$ that satisfy \eqref{eq:nl} for almost all times $t \in [0,T]$. We assume that the map $f$ satisfies certain regularity properties such that global existence of solutions to \eqref{eq:nl} is guaranteed, see, e.g., \cite{Khalil02}. 

In the remainder of this paper, it is assumed that the nonlinear system \eqref{eq:nl} is $\mathcal{L}_2$-stable in the sense that inputs $u \in \mathcal{L}_2$ are mapped to outputs and states $x, y \in \mathcal{L}_2$.


\subsection{Problem formulation}
We are primarily interested in characterizing input-output properties of \eqref{eq:nl}. In particular, we study a form of \emph{mixed dissipativity}, as characterized in the following definition.

\begin{definition}\label{def:msg}
    We say that the nonlinear system \eqref{eq:nl} possesses a mixed dissipativity property if there exist matrices $\Theta, \Pi \in \mathbb{S}^{2}$ and another matrix $\Psi_\varepsilon := \begin{bmatrix}\begin{smallmatrix}
2\varepsilon & 1 \\
1 &0 \end{smallmatrix}\end{bmatrix}$ with some $\varepsilon \in \mathbb{R}$ such that each input-output pair $\xi = [u,y]^\top$ of \eqref{eq:nl} satisfies: 
\begin{equation}\label{eq:gain1}
   0 \leq \langle \xi, \Theta \xi\rangle, 
\end{equation}
and / or:
   \begin{equation} \label{eq:gain2}
    0 \leq \langle \xi, \Psi_\varepsilon \xi \rangle, \quad \textup{and} \quad 0 \leq \langle \xi, \Pi \xi\rangle. 
    \end{equation}
   We call the system finite-gain mixed dissipative if 
\begin{align}
   \Theta_{11} \geq 0, \:\Theta_{22} <0, \:\textup{ and }\:\Pi_{11} \geq 0, \:\Pi_{22} <0.
    \end{align}
\end{definition}

Unlike \cite{Griggs07, Griggs09, Griggs11, Forbes10, Forbes13, Forbes13_2}, where mixed systems are described in terms of frequency-dependent supply rates, the characterization in Definition~\ref{def:msg} hinges on splitting the space of input signals into those signals for which the system in \eqref{eq:nl} satisfies \eqref{eq:gain1}, or \eqref{eq:gain2}, or both. This avoids the need for working with frequency-dependent supply rates directly, and allows for developing algorithmic methods to assess mixedness properties, as we will show. 

To develop some further intuition for the characterization in Definition~\ref{def:msg}, consider matrices of the form
\begin{equation}\label{eq:SMGM}
    \Theta = {\small\begin{bmatrix}
        \mu^2 & 0 \\ 
        0 & -1
    \end{bmatrix}}, \quad \Pi = {\small\begin{bmatrix}
       \gamma^2 & 0 \\
        0 & -1
    \end{bmatrix}}, \quad \textup{and} \quad \varepsilon = 0.
\end{equation}
Then, for some inputs the system exhibits finite-gain behaviour characterized by \eqref{eq:gain1}, i.e., the system satisfies $\|y\| \leq \mu \|u\|$, whereas for other inputs the system is finite-gain passive as characterized by the inequalities in \eqref{eq:gain2}, i.e., for these inputs the system satisfies both $0 \leq \langle u,y\rangle$ and $\|y\|\leq \gamma \|u\|$. 
{The system is therefore $\mathcal{L}_2$-stable in this case.}
Allowing $\Theta$ and $\Pi$ to be arbitrary symmetric matrices extends the idea of mixed passivity/finite-gain to include more general dissipativity properties beyond mixed passivity/finite-gain. 



The main objective of this paper is to characterize mixed input-output properties through the framework of dissipative systems. For this purpose, we draw inspiration from the generalized KYP lemma for LTI systems.   

\subsection{The generalized KYP lemma for LTI systems}\label{sec:prelim}
Consider an LTI system of the form 
\begin{equation}\label{eq:AB}
\begin{split}
\dot{x}(t) &= Ax(t)+Bu(t), \quad x(0) = 0, \\
y(t) & = Cx(t)+Du(t),
\end{split}
\end{equation}
with $x(t) \in \mathbb{R}^n$ the state, and $u(t), y(t) \in \mathbb{R}$ the input and output at times $t \in \mathbb{R}_{\geq 0}$. The generalized KYP lemma establishes equivalences between dissipativity of \eqref{eq:AB}, and input-output properties of \eqref{eq:AB}, expressed in the time domain, that only hold true for \emph{specific inputs}.

\begin{lemma}[\cite{Iwasaki05}]\label{lem:gkyp}
Consider the LTI system in \eqref{eq:AB} and let a matrix $\Theta \in \mathbb{S}^2$, and a real parameter $\bar{\omega}$ be given. Assume that $A$ is Hurwitz and \eqref{eq:AB} is controllable. Then, the following statements are equivalent\footnote{{Equivalence is not explicitly stated in \cite{Iwasaki05}, but follows directly from realizing that item 1) is equivalent to the LMI conditions appearing in \cite[Theorem 1]{Iwasaki05} and combining this result with \cite[Theorem 3]{Iwasaki05}.}}:
\begin{enumerate}
\item There exist functions $V(x) = x^\top P x$ and $W(x, \dot{x}) = \dot{x}^\top Q\dot{x}-\bar{\omega}^2 x^\top Q x$, with $Q \succ 0$ satisfying
\begin{equation}\label{eq:diss}
    \dot{V}(x)  \leq  W(x,\dot{x}) + \xi^\top \Theta \xi,
\end{equation}
where $\xi = [u,y]^\top$.
\item The time-domain inequality
\begin{equation}\label{eq:TDI}
0 \leq \int_{0}^{\infty} \xi(t)^\top \Theta \xi(t)dt  
\end{equation}
holds for all solutions of \eqref{eq:AB} with $u \in \mathcal{L}_2$ such that
\begin{equation}\label{eq:TDu}
\int_{0}^\infty \dot{x}(t)^\top Q \dot{x}(t)dt\leq \bar{\omega}^2 \int_{0}^{\infty} x(t)^\top Q x(t) dt.
\end{equation}
\end{enumerate}
\end{lemma}

The input dependent nature of the time-domain property in \eqref{eq:TDI} is embedded in \eqref{eq:diss} through the function $W$. Indeed, when integrating \eqref{eq:diss} from $t=0$ to $t = \infty$, and evaluating the term $\int_{0}^\infty W(x,\dot{x})dt$ in the right-hand side there are just two options: either this term is non-negative, or it is non-positive. The sign depends on the behaviour of $x$ and $\dot{x}$ over the whole time axis, forced by the input $u$. Hence, some inputs result in the term being non-negative, while the remaining inputs result in the term being non-positive. For those inputs which guarantee the latter (which is equivalent to \eqref{eq:TDu}), the time-domain inequality in \eqref{eq:TDI} holds true. When $W$ is negative semi-definite, \eqref{eq:TDI} holds true for all inputs and we recover the classical dissipativity result.

Next, we generalize these ideas to nonlinear systems.


\section{Main results}\label{sec:main}
This section presents the main results of this paper. In Theorem~\ref{th:0}, we present a dissipation-like characterization of the mixed dissipativity property. In Theorem~\ref{th:1}, we use this characterization for feedback stability analysis.   

\subsection{Characterization through storage functions}
\begin{theorem}\label{th:0}
    The system in \eqref{eq:nl} is mixed dissipative if there exist matrices $\Theta, \Pi \in \mathbb{S}^2$, a constant $\varepsilon \in \mathbb{R}$, locally Lipschitz continuous functions $S_i(x)$, $i = \left\{1,2,3\right\}$, with $S_i(0) = 0$, and locally Lipschitz functions $U(x,\dot{x})$, and $V(x,\dot{x})$ that satisfy \begin{subequations}\label{eq:VV}
\begin{align} 
    \dot{S}_{1}(x) &\leq U(x,\dot{x})+\xi^\top \Theta\xi, \label{eq:1} \\
    \dot{S}_{2}(x) &\leq V(x,\dot{x})+\xi^\top \Pi \xi, \label{eq:2}\\
    \dot{S}_{3}(x) &\leq V(x,\dot{x})+\xi^\top \Psi_\varepsilon \xi,\label{eq:3}\end{align}
\end{subequations}
with $\xi = [u,y]^\top$, and for any $(\dot{x}, x, u) \in \mathcal{L}_2$ satisfying \eqref{eq:nl}, at least one of either terms $\int_{0}^\infty U(x(t), \dot{x}(t))dt$ or  $\int_{0}^\infty V(x(t), \dot{x}(t))dt$ is non-positive.  
\end{theorem}

\begin{proof}
Integrating the conditions in \eqref{eq:VV} from $t = 0$ to $t = T$ and using $x(0)=0$ we find
\begin{align}\label{eq:S1}
   S_1(x(T)) & \leq \langle\xi, \Theta \xi\rangle_T+\int_{0}^T U(x(t),\dot{x}(t)dt, 
    \end{align}
    and
        \begin{subequations}\label{eq:S2}
        \begin{align}
         S_2(x(T))  & \leq \langle \xi, \Pi \xi\rangle_T + \int_{0}^T V(x(t),\dot{x}(t)dt, \\
         S_3(x(T)) & \leq \langle \xi, \Psi_\varepsilon \xi\rangle_T + \int_{0}^T V(x(t),\dot{x}(t)dt. 
        \end{align}
    \end{subequations}

Under the assumption that solutions to \eqref{eq:nl} are absolutely continuous and square-integrable, it follows from Barbalat's lemma \cite[Lemma 8.2]{Khalil02} that $\lim_{t\to\infty}x(t) = 0$. Hence, letting $T \to~\infty$ the functions $S_1$, $S_2$, and $S_3$ in the left-hand sides of \eqref{eq:S1} and \eqref{eq:S2} vanish. Under the hypothesis of the theorem, for every $(\dot{x},x,u) \in \mathcal{L}_2$ that satisfy \eqref{eq:nl} we have $\int_{0}^\infty U(x(t), \dot{x}(t))dt\leq 0$ and/or $\int_{0}^\infty V(x(t), \dot{x}(t))dt\leq 0$. Hence, for every trajectory of \eqref{eq:nl}, forced by the input $u$, either one or both of the inequalities in \eqref{eq:gain1} and \eqref{eq:gain2} are true, {as implied by \eqref{eq:S1} and \eqref{eq:S2}}. This completes the proof.
\end{proof}

Similar to the generalized KYP lemma (Lemma~\ref{lem:gkyp}), we do not require the {storage functions $S_i$, $i=\left\{1,2,3\right\}$ to be non-negative since, under the assumption that solutions to \eqref{eq:nl} are absolutely continuous and square-integrable, these functions vanish for $T \to \infty$. If the storage functions are taken to be non-positive, we can drop the aforementioned assumptions, but this restricts the class of functions we may consider. In the literature, integral inequalities that only hold for $T = \infty$ are often referred to as ``soft'', whereas inequalities that hold for any $T\geq 0$ are known as ``hard'' inequalities \cite{Megretski}.} 

The input-dependent nature of the input-output properties is embedded through the functions $U$ and $V$ in \eqref{eq:VV}. At this point, good choices for these functions are not obvious. However, for specific classes of nonlinear systems such as piecewise linear systems, a sensible choice would be based on piecewise quadratic functions \cite{Eijnden24}. We provide inspiration for the construction of possible functions in the next example.

\begin{example}[Nonlinear system]
Consider the system
\begin{equation}\label{eq:sys}
    \begin{split}
\dot{x} & = -x + u \\
y & = \varphi(x),
    \end{split} \quad \textup{ where } \quad \varphi(x) = \begin{cases}
x & \textup{ if } x \geq 0, \\
-\alpha x & \textup{ if } x \leq 0,        
    \end{cases}
\end{equation}
and with $\alpha \in (0,1)$. For positive inputs $u$ (and zero initial conditions), the state $x$ is positive, and $y = x$ for all times, i.e., we end up with a passive linear system. For negative inputs, the state $x$ is negative for all times, and thus $y = -\alpha x$ for all times. Hence, we end up with a linear system that violates passivity, but does admit a finite gain of $\alpha$. These observations hint toward the mixed input-output nature of the system. To formally show that the system is mixed, we will construct functions that satisfy the conditions of Theorem~\ref{th:0}.

First, consider the candidate function
\begin{equation}\label{eq:Vgamma}
   {S}_1(x) = x^2/\epsilon, \quad 0 < \epsilon < 1,
\end{equation}
for which the time-derivative satisfies
\begin{equation}\label{eq:DS1}
    \dot{S}_1(x) = -2x^2/\epsilon+2xu/\epsilon.
\end{equation}
For the system in \eqref{eq:sys} we have the following identities:
\begin{subequations}
    \begin{align}
        \varphi(x)x - x^2 & = 0 \textup{ if } x \geq 0, \\
        \varphi(x)x + \alpha x^2 & = 0 \textup{ if } x \leq 0.
    \end{align}
\end{subequations}
Using the above identities along with the facts that $y =x$ if $x \geq 0$ and $y = -\alpha x$ if $x \leq 0$, and by using Young's inequality we can upper-bound \eqref{eq:DS1} by 
\begin{equation*}
    \dot{S}_1(x) \leq \begin{cases}
        -\frac{1}{\epsilon}y^2 + \frac{1}{\epsilon}u^2 + c(\varphi(x)x-y^2) &\textup{ if } x \geq 0,\\
        -\frac{1}{\epsilon \alpha^2}y^2 + \frac{1}{\epsilon}u^2 + c(\varphi(x)x+\frac{1}{\alpha}y^2) &\textup{ if } x < 0.\\
    \end{cases}
\end{equation*}
Choosing $c = -1$ yields the common upper-bound 
\begin{equation}
    \dot{S}_1(x) \leq \left(\frac{\epsilon-1}{\epsilon} \right)y^2 + \frac{1}{\epsilon} u^2-\varphi(x)x, 
\end{equation}
and we select $V(x,\dot{x}) := -\varphi(x)x$. Next, consider \begin{equation}\label{eq:Vmu}
    S_2 (x) = \alpha x^2.
\end{equation}
Similar as before, we find an upper-bound on the time-derivative to be given by
\begin{equation*}
    \dot{S}_2(x) \leq \begin{cases}
        -\alpha y^2 + \alpha u^2 + k(\varphi(x)x-y^2) &\textup{ if } x \geq 0,\\
        -\frac{1}{\alpha} y^2 + \alpha u^2+ k(\varphi(x)x+\frac{1}{\alpha}y^2) &\textup{ if } x < 0,\\
    \end{cases}
\end{equation*}
where $k \in \mathbb{R}$. Choosing $k = \frac{1-\alpha^2}{1+\alpha}$ yields
\begin{equation}\label{eq:S2d}
    \dot{S}_2(x) \leq -y^2 + \alpha u^2 + k \varphi(x)x.
\end{equation}
To improve our estimates we add the term $\delta uy - \delta \varphi(x)(\dot{x}+x) = 0$ with $\delta<k$ to \eqref{eq:S2d} to obtain
\begin{equation}
    \dot{S}_2(x) \leq-y^2 + \delta uy+ \alpha u^2 + k \varphi(x)x -\delta \varphi(x)(\dot{x}+x),
\end{equation}
and select $U(x,\dot{x}) := k \varphi(x)x-\delta\varphi(x)(\dot{x}+x)$. Finally, consider the function
\begin{equation}\label{eq:Vepsilon}
    S_3 (x) = \int_{0}^x \varphi(s)ds.
\end{equation}
The time-derivative of this function satisfies
\begin{equation}\label{eq:dS3}
    \dot{S}_3 (x) = \varphi(x) \dot{x} = uy - \varphi(x)x = uy + V(x,\dot{x}).
\end{equation}

To show that $U(x,\dot{x})$ and $V(x,\dot{x})$ defined above satisfy the conditions of Theorem~\ref{th:0}, consider the integral terms
\begin{equation}\label{eq:intU}
    \int_{0}^\infty U(x,\dot{x})dt = \int_{0}^\infty (k\varphi(x)x-\delta\varphi(x)(\dot{x}+x))dt
\end{equation}
and 
\begin{equation}\label{eq:intV}
    \int_{0}^\infty V(x,\dot{x}) dt = -\int_{0}^\infty \varphi(x)x dt.
\end{equation}
We need to show that for each $(\dot{x},x,u)$ satisfying \eqref{eq:sys}, at least one of the integral terms is non-positive. If \eqref{eq:intV} is non-positive, there is nothing to show in addition. On the other hand, if \eqref{eq:intV} is non-negative, we need to show that \eqref{eq:intU} is non-positive. To do so, first note that by integrating \eqref{eq:dS3} from $t = 0$ to $t = \infty$ we find that in all cases 
\begin{equation}
  \int_{0}^\infty \varphi(x)(\dot{x}+x)dt=  \int_{0}^\infty uy dt \geq  \int_{0}^\infty \varphi(x) x dt.
\end{equation}
It then follows that 
\begin{equation}
\begin{split}
    \int_{0}^\infty U(x,\dot{x}) dt &\leq (k-\delta) \int_0^\infty \varphi(x)x dt.
    \end{split}
\end{equation}
Since $\delta < k$, and under the assumption that $\int_{0}^\infty \varphi(x)x dt \leq 0$ it follows that \eqref{eq:intU} is non-positive in this case. Hence, in all cases at least one of \eqref{eq:intU} or \eqref{eq:intV} is non-positive, and all conditions of Theorem~\ref{th:0} are satisfied. As such, the system is mixed dissipative with $\varepsilon = 0$, 
\begin{equation}\label{eq:TPnl}
    \Theta = {\small\begin{bmatrix}
        \alpha & \frac{\delta}{2} \\
        \frac{\delta}{2} & -1
    \end{bmatrix}}, \:\:\textup{ and }\:\: \Pi = {\small\begin{bmatrix}
        1 & 0 \\
        0 & -1
    \end{bmatrix}}.
\end{equation}
\end{example}

The choices for the functions $S_1$ and $S_2$ in \eqref{eq:Vgamma} and \eqref{eq:Vmu} result from the classical storage functions for asserting gain properties of a (non)linear system. The function in \eqref{eq:Vepsilon} is often used for passive nonlinear systems, see, for instance, \cite[Chapter 6]{Khalil02}. The term $\int_{0}^\infty \varphi(x(t))x(t)dt$ essentially ``averages'' the effect of switching between outputs $y = x$ and $y = -\alpha x$. That is, if  $\int_{0}^\infty \varphi(x(t))x(t)dt \geq 0$, the case where $\varphi(x) = x$ has the largest contribution, such that the system overall behaves as a passive system. On the other hand, if $\int_{0}^\infty \varphi(x(t))x(t)dt \leq 0$, the case where $\varphi(x) = -\alpha x$ contributes most, such that overall the system behaves as a system with small-gain. It must be mentioned that different (possibly better) choices for the functions $U$ and $V$ exist, and tighter estimates possibly could be obtained. In this regard, the above example merely aims at illustrating possible ways for constructing the functions involved in Theorem~\ref{th:0}, rather than providing the tightest dissipativity estimates.  

\subsection{An interconnection result}
\begin{figure}[htbt!]
        \centering
            \includegraphics[width=.3\textwidth]{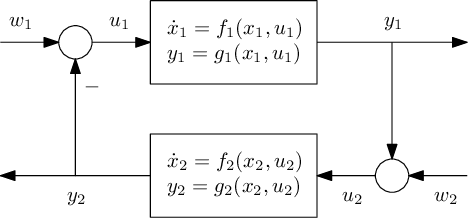}
        \caption{Feedback interconnection.}
        \label{fig:FB}
    \end{figure}

Consider the feedback interconnection shown in Fig.~\ref{fig:FB}, where $w_1,w_2 \in \mathcal{L}_2$ are external inputs. This system is said to be well-posed if, given $w_1, w_2 \in \mathcal{L}_{2e}$, there exist $y_1, y_2 \in \mathcal{L}_{2e}$ depending causally on $w_1, w_2$. 

The next result provides conditions for input-output stability of the interconnection using mixed properties. 

\begin{theorem}\label{th:1}
    Consider the feedback interconnection in Fig.~\ref{fig:FB} and suppose that subsystem $1$ is finite-gain mixed dissipative with $\varepsilon_1 \leq 0$, and subsystem $2$ is mixed dissipative with $\varepsilon_2 < 0$. Assume that the feedback interconnection is well-posed when $y_2$ is multiplied by any gain $\tau \in [0, 1]$.  Then, the interconnection maps $u \in \mathcal{L}_2$ to $ y \in \mathcal{L}_2$, and there exists $\gamma > 0$ such that $\norm{y} \leq \gamma \norm{u}$, if there exist constants $p_i \geq 0$, $i = \left\{1,2,3\right\}$ such that 
\begin{subequations}\label{eq:cond}
\begin{align}
    M^\top \Theta_1 M+ p_1 \Pi_2 & \prec 0,\label{eq:TT00} \\
     M^\top\Pi_1 M + p_2 \Theta_2  &\prec 0, \label{eq:TT01}\\
     M^\top \Theta_1 M + p_3 \Theta_2 & \prec 0, \label{eq:TT12} 
\end{align}
\end{subequations}
where $M = \begin{bmatrix}
    \begin{smallmatrix}
        0 & -1 \\ 1 & 0
    \end{smallmatrix}
\end{bmatrix}$. 
\end{theorem}

A direct proof of Theorem~\ref{th:1} is given in Appendix~\ref{app:pf2}. 

When the matrices $\Theta_i, \Pi_i$, $i=\left\{1,2\right\}$ in \eqref{eq:cond} are diagonal matrices of a form similar to \eqref{eq:SMGM}, Theorem~\ref{th:1} reduces to the rolled-off passivity theorem \cite[Theorem 1]{Chaffey22}. {We recover the classical passivity theorem by setting $\mu_i = 0$, and letting $\gamma_i \to \infty$, $i=\left\{1,2\right\}$, and for $\gamma_i<\mu_j$, $i,j = \left\{1,2\right\}$, $i\neq j$ we recover the small-gain theorem for operators on $\mathcal{L}_2$.}

We conclude this section with an interconnection example. 

\begin{example}[Feedback]
    Consider the nonlinear system in \eqref{eq:sys} (subsystem $1$), now placed in negative feedback interconnection with an LTI system $G$ (subsystem $2$) given by {
    \begin{equation}\label{eq:G}
        G(s) = \frac{3}{(s+1)(s+2)}.
    \end{equation}
    We verify through the generalized KYP-lemma that this LTI system is finite-gain mixed passive with $\varepsilon_2<0$, and $\Theta_2, \Pi_2$ as in \eqref{eq:SMGM} with $\mu_2 = 0.7$ and $\gamma_2 = 1.51$.} We will verify the conditions in Theorem~\ref{th:1} with $\Theta_1, \Pi_1$ as in \eqref{eq:TPnl}. Solving the LMIs in \eqref{eq:cond} over a range for $\alpha \in (0, 1)$ we find that the LMIs are feasible for all $\alpha < 0.43$. {Stability guarantees could not be given through the classical small-gain theorem (which requires $\gamma_1 \gamma_2 <1$, but in this example $\gamma_1\gamma_2 = 1.51$), nor through the passivity theorem as neither \eqref{eq:sys} nor \eqref{eq:G} are passive, thereby demonstrating the merit of our results.}  
\end{example}

\section{Conclusions}\label{sec:conclusions}
In this paper, we have provided an approach for characterizing nonlinear systems with mixed input-output properties through a dissipativity-like framework. Our results draw inspiration from the generalized KYP lemma for LTI systems and show that a system is mixed if it is dissipative with respect to a specialized supply rate, which encodes the input-dependent nature of the system behavior. We used the mixed characterizations to formulate a feedback stability result that relaxes classical small-gain and passivity results.  

One of the applications that may benefit from the ideas put forward in this paper comes from estimating the recently introduced scaled (relative) graph of a nonlinear system. For future work, we aim at tightening these estimates, and establishing further connections with the closely related framework of integral quadratic constraints. 



\appendix
\subsection{Proof of Theorem~\ref{th:1}}\label{app:pf2}
{
To show that the closed-loop systems maps $\mathcal{L}_2$ to $\mathcal{L}_2$, we will exploit a homotopy argument.  Following \cite[Thm. 3.2]{Freeman2022}, we place a gain $\tau \in [0,1]$ in the loop after $y_2$, and show that, for each $\tau$ and any inputs $w_1, w_2 \in \mathcal{L}_2$ such that the internal signals $u_1, u_2, y_1, y_2 \in \mathcal{L}_2$, the input-output gain is finite.} 
{
As such, let $\tau \in [0, 1]$ and assume that $w_1, w_2 \in \mathcal{L}_2$ generate $u_1, u_2, y_1, y_2 \in \mathcal{L}_2$.  We can write}
\begin{equation*}
\begin{split}
\begin{bmatrix}
    u_1 \\ y_1
\end{bmatrix} &= \begin{bmatrix}
    0 & -\tau \\
    1 & 0
\end{bmatrix}\begin{bmatrix}
    y_1 \\ y_2
\end{bmatrix} + \begin{bmatrix}
    1 & 0 \\ 0 & 0
\end{bmatrix}\begin{bmatrix}
    w_1 \\ w_2
\end{bmatrix} = M_\tau y + N_1w,
\end{split}
\end{equation*}
and
\begin{equation*}
\begin{split}
\begin{bmatrix}
    u_2 \\ y_2
\end{bmatrix} &= \begin{bmatrix}
    1 & 0 \\
    0 & 1
\end{bmatrix}\begin{bmatrix}
    y_1 \\ y_2
\end{bmatrix} + \begin{bmatrix}
    0 & 1 \\ 0 & 0
\end{bmatrix}\begin{bmatrix}
    w_1 \\ w_2
\end{bmatrix} = y + N_2w.
\end{split}
\end{equation*}
Suppose that both subsystems are excited in a manner that they both exhibit dissipative behaviour characterized by $\Theta_1$ and $\Theta_2$, i.e, both satisfy \eqref{eq:gain1}. Then, we can write
\begin{equation}\label{eq:pf1}
0 \leq \left\langle \begin{bmatrix}
    y \\ w
\end{bmatrix}, \begin{bmatrix}
    M_\tau ^\top \Theta_1 M_\tau & N_1^\top \Theta_1 M_\tau \\
    M_\tau^\top \Theta_1 N_1 & N_1^\top \Theta_1 N_1
\end{bmatrix}\begin{bmatrix}
    y \\ w
\end{bmatrix}\right\rangle
\end{equation}
for the first subsystem. In a similar manner, we find
\begin{equation}\label{eq:pf2}
0 \leq \left\langle \begin{bmatrix}
    y \\ w
\end{bmatrix}, \begin{bmatrix}
    \Theta_2 & N_2^\top \Theta_2 \\
    \Theta_2 N_2 & N_2^\top \Theta_2 N_2
\end{bmatrix}\begin{bmatrix}
    y \\ w
\end{bmatrix}\right\rangle
\end{equation}
for the second subsystem. Multiplying inequality \eqref{eq:pf2} with a non-negative number $p_3(\tau)$ that may depend continuously on $\tau$, and adding the result to inequality \eqref{eq:pf1} yields
\begin{equation} \label{eq:complete}
    0 \leq \left\langle \begin{bmatrix}
    y \\ w
\end{bmatrix}, \begin{bmatrix}
    Q(\tau) & R(\tau) \\
    R^\top(\tau)  & S(\tau)
\end{bmatrix}\begin{bmatrix}
    y \\ w
\end{bmatrix}\right\rangle,
\end{equation}
with matrices 
\begin{subequations}
    \begin{align}
        Q(\tau) &= M_\tau ^\top \Theta_1 M_\tau+p_3(\tau)\Theta_2 \label{eq:Q}\\
        R(\tau)&=N_1^\top \Theta_1 M_\tau+p_3(\tau)N_2^\top \Theta_2 \\
        S(\tau)&=N_1^\top \Theta_1 N_1+p_3(\tau)N_2^\top \Theta_2 N_2.
    \end{align}
\end{subequations}
We need to show that, under the hypothesis of the theorem, the matrix $Q(\tau)$ in \eqref{eq:Q} is negative definite for all $\tau \in [0,1]$. For this purpose, let us denote
\begin{equation}
    \Theta_i = {\small\begin{bmatrix}
        a_i & b_i \\
        b_i & c_i
    \end{bmatrix}}
\end{equation}
and expand $Q(\tau)$ as
\begin{equation}\label{eq:36}
    Q(\tau) = \begin{bmatrix}
        c_1& -\tau b_1 \\
        -\tau b_1 & \tau^2 a_1
    \end{bmatrix} + p_3(\tau) \begin{bmatrix}
        a_2 & b_2 \\
        b_2 & c_2
    \end{bmatrix}.
\end{equation}
Since subsystem $1$ is finite-gain mixed dissipative, we have $a_1 \geq 0$ and $c_1 <0$. Since $a_1\geq 0$, it follows from condition \eqref{eq:TT12} that $\min_{\tau \in [0,1]} p_3(\tau)=\tilde{p}_3 >0$ and $c_2 < -a_1/\tilde{p}_3 \leq 0$. Hence, when $\tau = 0$, we can choose $\tilde{p}_3$ sufficiently small, such that
\begin{equation}
    Q_0= \begin{bmatrix}
        c_1& 0 \\
        0 & 0
    \end{bmatrix} + \tilde{p}_3 \begin{bmatrix}
        a_2 & b_2 \\
        b_2 & c_2
    \end{bmatrix} \prec 0.
\end{equation}
Feasibility of \eqref{eq:TT12} implies that for $\tau = 1$, the matrix $Q(1):= Q_1$ is negative definite, i.e., $Q_1 \prec 0$. Taking the convex combination of $Q_0$ and $Q_1$, that is, $(1-\tau) Q_0+\tau Q_1$ with $\tau \in [0,1]$ yields
\begin{equation}\label{eq:M1}
   \begin{bmatrix}
        c_1 & -\tau b_1 \\
        -\tau b_1 & \tau a_1
    \end{bmatrix} + p_3(\tau) \begin{bmatrix}
        a_2 & b_2 \\b_2 & c_2
    \end{bmatrix}\prec 0,
\end{equation}
where $p_3(\tau) = (1-\tau) \tilde{p}_3+\tau p_3 >0$. Since $a_1\geq 0$ and $\tau \geq 0$, we find that $\tau^2 a_1 - \tau a_1 \leq 0$. Combining this with the inequality in \eqref{eq:M1} yields for all $\tau \in [0,1]$
\begin{equation}\label{eq:M2}
    Q(\tau) =  \begin{bmatrix}
        c_1 & -\tau b_1 \\
        -\tau b_1 & \tau^2 a_1
    \end{bmatrix} + p_3(\tau) \begin{bmatrix}
        a_2 & b_2 \\b_2 & c_2
    \end{bmatrix}\prec 0.
\end{equation}
Let $\epsilon~= \min_{\tau \in [0,1]} \left\{\lambda_1(-Q(\tau)), \lambda_2(-Q(\tau))\right\} > 0$, $r = \max_{\tau \in [0,1]} \|R(\tau)\|_2 \geq 0$, and $s = \max_{\tau \in [0,1]} \|S(\tau)\|_2 \geq 0$. Then, from \eqref{eq:complete} we find
\begin{equation*}
\begin{split}
    0 & \leq -\epsilon \|y\|^2 + r \|y\|\|w\| + s \|w\|^2 \leq \frac{\gamma^2}{2\epsilon}\|w\|^2 - \frac{\epsilon}{2}\|y\|^2,
    \end{split}
\end{equation*}
with $\gamma^2 = r^2+2\epsilon s$, resulting in $\|y\| \leq(\gamma/\epsilon) \|w\|$. 

In a similar manner, we can find a finite input-output gain bound for all $\tau~\in [0,1]$ whenever subsystem $i$ satisfies \eqref{eq:gain1} with $\Theta_i$ and subsystem $j$ satisfies \eqref{eq:gain2} with $\Pi_j$ and $\Psi_{\varepsilon_j}$. 

When both subsystems are dissipative in the sense that they both satisfy \eqref{eq:gain2} with $\Pi_i$ and $\Psi_{\varepsilon_i}$, we arrive at the following inequalities. For subsystem $1$, note that since $\varepsilon_1 \leq  0$, the inequality $0 \leq \langle u_1,y_1\rangle$, also holds true. Then, we find
\begin{align}\label{eq:O1}
    0 \leq \left\langle \begin{bmatrix}
        y \\ w
    \end{bmatrix}, \begin{bmatrix}
        M_\tau^\top \Omega_1 M_\tau & N_1^\top \Omega_1 M_\tau \\
        M_\tau^\top \Omega_1 N_1 & N_1^\top \Omega_1 N_1    \end{bmatrix}\begin{bmatrix}
        y\\w
    \end{bmatrix}\right\rangle,
\end{align}
where $\Omega_1 = \alpha_1\Pi_1+\beta_1 \Psi_{0}$ with $\Psi_0 = \begin{bmatrix}
    \begin{smallmatrix}
        0 & 1\\ 1 &0
    \end{smallmatrix}
\end{bmatrix}$, and $\alpha_1, \beta_1 \geq 0$. Similarly, for the second subsystem we have
\begin{align}\label{eq:O2}
    0 \leq \left\langle \begin{bmatrix}
        y \\ w
    \end{bmatrix}, \begin{bmatrix}
        \Omega_2 & N_2^\top \Omega_2 \\
        \Omega_2 N_2 & N_2^\top \Omega_2 N_2  \end{bmatrix}\begin{bmatrix}
        y\\w
    \end{bmatrix}\right\rangle,
\end{align}
where $\Omega_2 = \alpha_2\Pi_2+\beta_2 \Psi_{\varepsilon_2}$ and $\alpha_2, \beta_2 \geq 0$. When adding \eqref{eq:O1} and \eqref{eq:O2}, we find
\begin{equation}\label{eq:LKH}
    0 \leq \left\langle \begin{bmatrix}
        y \\ w
    \end{bmatrix}, \begin{bmatrix}
        L(\tau) & K(\tau) \\
        K^\top(\tau) & H(\tau)
    \end{bmatrix}\begin{bmatrix}
        y \\ w
    \end{bmatrix}\right\rangle,
\end{equation}
where
\begin{subequations}
    \begin{align}
        L(\tau) &= M_\tau ^\top \Omega_1 M_\tau+\Omega_2\label{eq:L1app}\\
        K(\tau)&=N_1^\top \Omega_1 M_\tau+N_2^\top \Omega_2 \\
        H(\tau)&=N_1^\top \Omega_1 N_1+N_2^\top \Omega_2 N_2.
    \end{align}
\end{subequations}
We need to show that for all $\tau \in [0,1]$, the matrix $L(\tau)$ in \eqref{eq:L1app} is negative definite. To do so, let us write $\Pi_i$ as
\begin{equation}
    \Pi_i = {\small\begin{bmatrix}
        \mu_i & \sigma_i \\
        \sigma_i&  \rho_i
    \end{bmatrix}}
\end{equation}
and expand $L(\tau)$ as $L(\tau)=\mathcal{L}_1(\tau)+\mathcal{L}_2(\tau)$, with
\begin{subequations}
\begin{align}
    \mathcal{L}_1(\tau) & = {\small\begin{bmatrix}
       \alpha_1 \rho_1 & -\tau (\alpha_1\sigma_1+\beta_1) \\
        -\tau(\alpha_1\sigma_1+\beta_1) & \tau^2(\alpha_1 \mu_1)
    \end{bmatrix}} \\
    \mathcal{L}_2(\tau) & = {\small\begin{bmatrix}
        \alpha_2 \mu_2 + 2 \beta_2 \varepsilon_2 & \alpha_2 \sigma_2 + \beta_2\\
        \alpha_2 \sigma_2 + \beta_2 & \alpha_2 \rho_2 
    \end{bmatrix}}.
\end{align}
\end{subequations}
We let $\alpha_i, \beta_i$, $i=\left\{1,2\right\}$ depend continuously on $\tau$, i.e., for each $\tau$ we can possibly choose a different $\alpha_i, \beta_i$.
Consider $\tau = 0$. In this case $L(0):=L_0$ reduces to
\begin{equation}
    L_0 = {\small\begin{bmatrix}
       \tilde{\alpha}_1 \rho_1+\tilde{\alpha}_2 \mu_2 + 2\tilde{\beta}_2 \varepsilon_2 & \tilde{\alpha}_2 \sigma_2 + \tilde{\beta}_2\\
        \tilde{\alpha}_2 \sigma_2 + \tilde{\beta}_2 & \tilde{\alpha}_2 \rho_2  
    \end{bmatrix}}.
\end{equation}
Since the first subsystem is finite-gain mixed dissipative, we have $\mu_1 \geq 0$ and $\rho_1 <0$. Moreover, condition \eqref{eq:TT01} implies (through a similar argument as stated after \eqref{eq:36}) that $\rho_2 <0$. Setting $\tilde{\alpha}_1 = \alpha_1$ we can always choose $\tilde{\alpha}_2>0$ and $\tilde{\beta}_2>0$ sufficiently small to guarantee $L_0 \prec 0$. 

Next, consider $\tau = 1$. In this case $L(1) :=L_1$ is given by
\begin{equation*}
    L_1 = {\small\begin{bmatrix}
        \alpha_1 \rho_1 + \alpha_2 \mu_2 +2\beta_2 \varepsilon_2 & \alpha_2\sigma_2-\alpha_1\sigma_1 +\beta_2-\beta_1 \\
        \alpha_2\sigma_2-\alpha_1\sigma_1+\beta_2-\beta_1 & \alpha_1\mu_1 + \alpha_2 \rho_2
    \end{bmatrix}}.
\end{equation*}
Take $\beta_1 = \beta_2 >0$, and $\alpha_1 = k \alpha_2$. Then, we can choose $k$ sufficiently small such that $k\mu_1 + \rho_2 <0$. Next, we can choose $\beta_2$ sufficiently large to guarantee that $L_1 \prec 0$. Taking the convex combination of $L_0$ and $L_1$, that is, $(1-\tau)L_0+\tau L_1$ with $\tau \in [0,1]$ yields $    \tilde{L}(\tau) = \tilde{\mathcal{L}}_1(\tau)+\tilde{\mathcal{L}}_2(\tau) \prec 0$, with
\begin{subequations}
    \begin{align}
        \tilde{\mathcal{L}}_1(\tau) &= {\small\begin{bmatrix}
        \alpha_1 \rho_1 & -\tau(\alpha_1\sigma_1+\beta_1) \\
       -\tau(\alpha_1\sigma_1+\beta_1) &
       \tau(\alpha_1 \mu_1)
    \end{bmatrix}}\label{eq:Ltile1}\\
     \tilde{\mathcal{L}}_2(\tau) & = {\small\begin{bmatrix}
        \alpha_2(\tau)\mu_2 + 2 \beta_2(\tau) \varepsilon_2 & \alpha_2(\tau)\sigma_2 +\beta_2(\tau)\\
       \alpha_2(\tau)\sigma_2 +\beta_2(\tau) &
       \alpha_2(\tau) \rho_2
    \end{bmatrix}}
    \end{align}
\end{subequations}
and where $\alpha_2(\tau) = (1-\tau)\tilde{\alpha}_2 + \tau \alpha_2$, and $\beta_2(\tau) = (1-\tau) \tilde{\beta}_2 + \tau \beta_2$. Since $\alpha_1 \mu_1 \geq 0$ we can replace the bottom right part in \eqref{eq:Ltile1} by $\tau^2\alpha_1 \mu_1$ to find $L(\tau) \preceq \tilde{L}(\tau) \prec 0$ for all $\tau \in [0,1]$. Let $\delta = \min_{\tau \in [0,1]}\left\{-\lambda_1(-L(\tau)),-\lambda_2(-L(\tau))\right\} >0$, $m = \max_{\tau \in [0,1]}\|K(\tau)\|_2\geq 0$, and $n = \max_{\tau \in [0,1]}\|H(\tau)\|_2 \geq 0$. Then, from \eqref{eq:LKH} we find 
\begin{equation*}
    0 \leq - \delta \|y\|^2 + m\|y\|\|w\| + n \|w\|^2 \leq \frac{\nu^2}{2\delta}\|w\|^2-\frac{\delta}{2}\|y\|^2,
\end{equation*}
with $\nu^2 = m^2+2\delta n$, leading to $\|y\| \leq ({\nu}/{\delta}) \|w\|$.

For each $\tau \in [0,1]$ {and $w,y \in \mathcal{L}_2$}, we find a finite gain-bound of the form $\|y\| \leq r_m \|w\|$. We furthermore note that the interconnection is assumed to be well-posed for all $\tau \in [0, 1]$.  It follows by homotopy \cite[Thm. 3.2]{Freeman2022} that the closed loop is finite gain stable for $\tau = 1$. This completes the proof.

\vspace{-0.01cm}
\bibliographystyle{IEEEtran}

\end{document}